\documentstyle[12pt,psfig,preprint,aps,tighten]{revtex}
\newcommand {\ignore}[1]{}

\def\gsim{\:\raisebox{-0.5ex}{$\stackrel{\textstyle>}{\sim}$}\:}

\def\21{$SU(2) \ot U(1)$}
\def\321{$SU(3) \ot SU(2) \ot U(1)$}

\def\ns{\hbox{$\nu_{s}$ }}

\def\n.c.#1#2#3{         { Nuovo Cim. }{\bf #1}, #3 (19#2)}
\def\r.n.c.#1#2#3{       { Riv. del Nuovo Cim. }{\bf #1}, #3 (19#2)}

\begin{document}
\draft
\preprint{\vbox{
\hbox{FTUV/00--13, IFIC/00--14}
} }
\title{Updated Global Analysis of the Atmospheric Neutrino Data \\
in terms of neutrino oscillations}

\author{
N.\ Fornengo$^{\mbox{a,b}}$
\footnote[1]{E-mail: fornengo@flamenco.ific.uv.es}, 
M.\ C.\ Gonzalez-Garcia$^{\mbox{a}}$
\footnote[4]{E-mail: concha@flamenco.ific.uv.es}, 
and J.\ W.\ F.\ Valle$^{\mbox{a}}$ 
\footnote[3]{E-mail: valle@flamenco.ific.uv.es}
}
\address{$^{\mbox{a}}$ Instituto de F\'{\i}sica Corpuscular - C.S.I.C. \\
 Departamento de F\'{\i}sica Te\`orica, Universitat de Val\`encia \\
Edificio Institutos de Paterna, Apt 2085, 46071 València, Spain \\
$^{\mbox{b}}$
Dipartimento di Fisica Teorica, Universit\`a di Torino \\
and INFN, Sez. di Torino, Via P. Giuria 1, I--10125 Torino, Italy}
\maketitle

\begin{abstract}
  A global analysis of all the available atmospheric neutrino data is
  presented in terms of neutrino oscillations in the $\nu_\mu \to
  \nu_\tau$ and $\nu_\mu \to \nu_s$ channels, where \ns denotes a
  sterile neutrino.  We perform our analysis of the contained events
  data as well as the upward--going neutrino--induced muon fluxes. In
  addition to the previous data samples of Frejus, Nusex, IMB and
  Kamioka experiments, we include the full data set of the 52 kton-yr
  of Super--Kamiokande, the recent 4.6 kton-yr contained events of
  Soudan2 and the results on upgoing muons from the MACRO and Baksan
  detectors.  From the statistical analysis it emerges that the
  $\nu_\mu \to \nu_\tau$ channel provides the best agreement with the
  combined data, with a best fit point of $\sin^2(2\theta) = 0.99$ and
  $\Delta m^2 = 3.0 \times 10^{-3}$ eV$^2$. Although somehow
  disfavoured, the $\nu_\mu \to \nu_s$ channels cannot be ruled out on
  the basis of the global fit to the full set of observables.

\end{abstract}


\section{Introduction}

Together with the solar neutrino problem \cite{Bahcall:1999rt} the
atmospheric neutrino anomaly constitutes the second evidence for
physics beyond the Standard Model. Indeed a large number of
observations
\cite{Frejus,Nusex,IMB,Soudan,Soudan2,Kamiokande,skcont,skup,sk99,MACRO,Baksan}
have been performed concerning electron-- and muon--neutrino fluxes
produced by hadronic showers initiated by cosmic--ray interactions in
the upper atmosphere. Apart from the first iron--calorimeter detectors
\cite{Frejus,Nusex}, all experiments, which also entail different
detection techniques, have steadily reported a deficit of the
collected number of events with respect to the theoretical
expectations \cite{fluxes1,fluxes2}.  Although the knowledge of the
fluxes of atmospheric neutrinos is affected by uncertainties which
range from about 20\% to 30\%, the expected ratio $R(\mu/e)$ of the
muon neutrino ($\nu_\mu + \bar{\nu}_\mu$) over the electron neutrino
flux ($\nu_e+\bar{\nu}_e$) is known with much better confidence, since
the uncertainties associated with the absolute fluxes largely cancel
out.
It is fair to ascribe an overall uncertainty less than about 5\% to
the calculated $R(\mu/e)$ ratio. Since this ratio is measured to be
substantially smaller than the expectations, one faces an anomaly
which only seems possible to account for in terms of non--standard
neutrino properties.

Super--Kamiokande high statistics observations~\cite{skcont,skup}
indicate that the deficit in the total ratio $R(\mu/e)$ is due to the
neutrinos arriving in the detector at large zenith angles. They also
show that the $e$-like events do not present any compelling evidence
of a zenith-angle dependent suppression, while the $\mu$-like event
rates are substantially suppressed at large angles. Different
explanations to these features have been proposed and discussed in the
literature \cite{others1,others2}.  The simplest and most direct
possibility is represented by the oscillation of muon neutrinos
$\nu_\mu$ into either a $\nu_\tau$ or a sterile neutrino $\nu_s$
\cite{atm98,atmo98,skos}.  The oscillation hypothesis provides a very
good explanation for this smaller-than-expected ratio, which is also
simple and well-motivated theoretically.

In this paper we present our updated analysis of all the available
data on atmospheric neutrinos in terms of neutrino oscillation. We
include in the analysis all the existing experimental results obtained
so far. In addition to the previous data samples of
Frejus\cite{Frejus}, Nusex\cite{Nusex}, IMB\cite{IMB} and
Kamioka\cite{Kamiokande} experiments, we also consider the recent
Soudan2\cite{Soudan2} data which refers to an exposure of 4.6 kton-yr,
the full data set of the 52 kton-yr of Super--Kamiokande\cite{sk99},
including both contained events and upgoing muon data, and finally the
results on upgoing muons reported from the MACRO\cite{MACRO} and
Baksan\cite{Baksan} experiments. We critically discuss the analysis of
the various individual data sets. Moreover we consider the combined
information that can be derived from all the experimental evidences
so--far obtained. We hope this will contribute to a more complete
understanding of the atmospheric neutrino anomaly in the framework of
a global analysis based on a self--consistent theoretical calculation
of the event rates. This has the advantage of calibrating all relevant
elements in the analysis, such as theoretical atmospheric neutrino
fluxes, in a consistent way. This allows different experiments to be
compared in a meaningful way. A new element which we now add to our
previous investigations\cite{atm98} is the inclusion of the upgoing
muon data and the corresponding theoretical flux
determinations\cite{sk99,MACRO,Baksan}. We anticipate that from the
statistical analysis the $\nu_\mu \to \nu_\tau$ emerges as the
oscillation channel which provides the best agreement with the
combined data. The $\nu_\mu \to \nu_s$ channels, although slightly
disfavoured, cannot be statistically ruled out on the basis of the
global fit to the full set of observables.

The outline of the paper is the following: in Sect. II we briefly
recall the theoretical calculation of the event rates for contained
events and upgoing muon fluxes, as well as the calculation of the
oscillation probabilities. Sect. III discusses the statistical
approach of our analysis and reports on the results of the fits. Sect.
IV presents a discussion of the results and the conclusions. We
include more details on the statistical analysis in an Appendix.

\section{Atmospheric Neutrino Induced Events in Underground Detectors}

Atmospheric neutrinos can be detected in underground experiments by
direct observation of their charged current interaction inside the
detector.  These are the so-called contained events, which can be
further classified into fully contained events when the charged lepton
(either electron or muon) produced by the neutrino interaction does
not escape the detector, and partially contained muons when the muon,
produced inside, leaves the detector. In the case of Kamiokande and
Super--Kamiokande, the contained data sample is further divided into
sub-GeV events with visible energy below 1.2 GeV and multi-GeV events
when the lepton energy is above such cutoff. On the average, sub-GeV
events arise from neutrinos of several hundreds of MeV while multi-GeV
events are originated by neutrinos with energies of the order of
several GeV.

Higher energy muon neutrinos and anti-neutrinos can also be detected
indirectly by observing the muons produced by charged current
interactions in the vicinity of the detector. These are the so called
upgoing muons. Should the muon stop inside the detector, it will be
classified as a ``stopping'' muon, while if the muon track crosses the
full detector the event is classified as a ``through-going'' muon.
Typically stopping muons arise from neutrinos of energies around ten
GeV while through-going muons are originated by neutrinos with
energies of the order of hundred GeV.

\subsection{Contained events}

At present, six underground experiments have collected data on
contained events.  Three of them, Kamiokande~\cite{Kamiokande}
IMB~\cite{IMB} and Super--Kamiokande~\cite{skcont} use water-Cerenkov
detectors, while the other three, Fr\'ejus~\cite{Frejus},
NUSEX~\cite{Nusex} and Soudan2~\cite{Soudan,Soudan2} are iron
calorimeter detectors.

For a given neutrino conversion mechanism, the expected number of
$\mu$-like and $e$-like contained events, $N_\alpha$, $\alpha = \mu,
e$ can be computed as:
\begin{equation}
N_\mu= N_{\mu\mu} +\
 N_{e\mu} \; ,  \;\;\;\;\;\
N_e= N_{ee} +  N_{\mu e} \; ,
\label{eventsnumber}
\end{equation}
where
\begin{eqnarray}
N_{\alpha\beta} &=& n_t T
\int
\frac{d^2\Phi_\alpha}{dE_\nu d(\cos\theta_\nu)} 
\kappa_\alpha(h,\cos\theta_\nu,E_\nu) 
P_{\alpha\beta} \frac{d\sigma}{dE_\beta}\varepsilon(E_\beta)
dE_\nu dE_\beta d(\cos\theta_\nu)dh\;
\label{event0}
\end{eqnarray}
and $P_{\alpha\beta}$ is the conversion probability of $\nu_\alpha \to
\nu_\beta$ for given values of $E_{\nu}, \cos\theta_\nu$ and $h$, {\em
  i.e.}, $P_{\alpha\beta} \equiv P(\nu_\alpha \to \nu_\beta; E_\nu,
\cos\theta_\nu, h ) $.  In the Standard Model (SM), the only non-zero
elements are the diagonal ones, {\em i.e.} $P_{\alpha\alpha}=1$ for
all $\alpha$.  In Eq.(\ref{event0}) $n_t$ denotes the number of
targets, $T$ is the experiment running time, $E_\nu$ is the neutrino
energy and $\Phi_\alpha$ is the flux of atmospheric neutrinos of type
$\alpha=\mu ,e$ for which we use the Bartol flux\cite{fluxes1};
$E_\beta$ is the final charged lepton energy and
$\varepsilon(E_\beta)$ is the detection efficiency for such charged
lepton; $\sigma$ is the neutrino-nucleon interaction cross section,
and $\theta_\nu$ is the angle between the vertical direction and the
incoming neutrinos ($\cos\theta_\nu$=1 corresponds to the down-coming
neutrinos).  In Eq.~(\ref{event0}), $h$ is the slant distance from the
production point to the sea level for $\alpha$-type neutrinos with
energy $E_\nu$ and zenith angle $\theta_\nu$. Finally, $\kappa_\alpha$
is the slant distance distribution, normalized to one
\cite{pathlength}. For the angular distribution of events we integrate
in the corresponding bins for $\cos\theta_\beta$ where $\theta_\beta$
is the angle of the detected lepton, taking into account the opening
angle between the neutrino and the charged lepton directions as
determined by the kinematics of the neutrino interaction.  In average
the angle between the directions of the final-state lepton and the
incoming neutrino ranges from $70^\circ$ at 200 MeV to $20^\circ$ at
1.5 GeV.

The neutrino fluxes, in particular in the sub-GeV range, depend on the
solar activity.  In order to take this fact into account, we use in
Eq.~(\ref{event0}) a linear combination of atmospheric neutrino fluxes
$\Phi_\alpha^{max}$ and $\Phi_\alpha^{min}$ which correspond to the
most active Sun (solar maximum) and quiet Sun (solar minimum),
respectively, with different weights which depend on the running
period of each experiment \cite{atm98}.
Following Ref.~\cite{atm98} we explicitly verify in our present
reanalysis the agreement of our predictions with the experimental
Monte Carlo predictions, leading to a good confidence in the
reliability of our results for contained events.

\subsection{Upward Going Muons}

Several experiments have obtained data on these neutrino induced
muons.  In our analysis we consider the results from three
experiments: Super--Kamiokande \cite{skup}, MACRO \cite{MACRO} and
Baksan\cite{Baksan}.  They present their upgoing muon data in the form
of measured muon fluxes.  We obtain the effective muon fluxes for both
stopping and through-going muons by convoluting the survival $\nu_\mu$
probabilities (calculated as in Sect. II.C) with the corresponding
muon fluxes produced by the neutrino interactions with the Earth. We
include the muon energy loss during propagation both in the rock and
in the detector according to Refs. \cite{muloss,ricardo} and we take
into account also the effective detector area for both types of
events, stopping and through-going.  Schematically
\begin{equation}
\Phi_\mu(\theta)_{S,T}=\frac{1}{A(L_{min},\theta)}
\int_{E_{\mu ,min}}^{\infty} 
\frac{d\Phi_\mu(E_\mu,\cos\theta)}{dE_\mu d\cos\theta}
A_{S,T}(E_\mu,\theta)dE_{\mu} \, \; ,
\label{upmuons1}
\end{equation}  
where
\begin{eqnarray}
\frac{d\Phi_\mu}{dE_\mu d\cos\theta}
&=& N_A \int_{E_{\mu}}^\infty dE_{\mu 0}
\int_{E_{\mu 0}}^\infty dE_\nu
\int_0^\infty dX  \int_0^\infty dh \;
\kappa_{\nu_\mu}(h,\cos\theta,E_\nu) \nonumber \\
& & \frac{d\Phi_{\nu_\mu}(E_\nu,\theta)}{dE_\nu d\cos\theta}
P_{\mu\mu}
\frac{d\sigma(E_\nu,E_{\mu 0})}{dE_{\mu 0}}\,
F_{rock}(E_{\mu 0}, E_\mu, X)
\label{upmuons2}
\end{eqnarray}
$N_A$ is the Avogadro number, $E_{\mu 0}$ is the energy of the muon
produced in the neutrino interaction and $E_\mu$ is the muon energy
when entering the detector after traveling a distance $X$ in the rock.
$\cos\theta$ labels both the neutrino and the muon directions which to
a very good approximation at the relevant energies are collinear.  We
denote by $F_{rock}(E_{\mu 0}, E_\mu, X)$ the function which
characterizes the energy spectrum of the muons arriving the detector.
Following standard practice \cite{muloss,ricardo} in our calculations
we use the analytical approximation obtained by neglecting the
fluctuations during muon propagation in the Earth.  In this case the
three quantities $E_{\mu 0}$, $E_\mu$, and $X$ are not independent:
\begin{equation}
\int_{0}^\infty F_{rock}(E_{\mu 0}, E_\mu, X) dX=
\frac{1}{\langle d{\cal E}_\mu(E_\mu)/dX\rangle} \; ,
\end{equation}
where $\langle d{\cal E}_\mu(E_\mu)/dX\rangle$ is the average muon
energy loss due to ionization, bremsstrahlung, $e^+e^-$ pair
production and nuclear interactions in the rock according to Refs.
\cite{muloss,ricardo}.  Equivalently, the pathlength traveled by the
muon inside the Super--Kamiokande detector is given by the muon range
function in water
\begin{equation}
\label{Emcut}
L(E_\mu)=\int_{0}^{E_\mu}
\frac{1}{\langle d{\cal E}_\mu(E'_\mu)/dX\rangle} \; dE'_\mu \; .
\end{equation} 
In Eq.(\ref{upmuons1})
$A(L_{min},\theta)=A_{S}(E_\mu,\theta)+A_{T}(E_\mu,\theta)$ is the
projected detector area for internal path-lengths longer than
$L_{min}$ which for Super--Kamiokande is $L_{min}$ = 7 m. Here $A_{S}$
and $A_{T}$ are the corresponding effective areas for stopping and
through-going muon trajectories.  For Super--Kamiokande we compute
these effective areas using the simple geometrical picture given in
Ref.~\cite{lipari1}. For a given angle, the threshold energy cut for
Super--Kamiokande muons is obtained by equating Eq.(\ref{Emcut}) to
$L_{min}$, i.e.  $L(E_{\mu}^{\mathrm th}) =L_{min}$.

In contrast with Super--Kamiokande, Baksan and MACRO present their
results as muon fluxes for $E_\mu>1$ GeV, after correcting for
detector acceptances. Therefore in this case we compute the expected
fluxes as Eqs.~(\ref{upmuons1}) and (\ref{upmuons2}) but without the
inclusion of the effective areas.

We have explicitly verified the agreement of our predictions for
upgoing muons with the experimental Monte Carlo predictions from
Super--Kamiokande, Baksan and MACRO. The agreement can be observed by
comparing our Standard Model predictions for the angular distributions
in Fig.~\ref{angup} with the corresponding distributions in
Refs.~\cite{sk99} and~\cite{MACRO}. We find an agreement to the 5\%
and 1\% level, respectively.

\subsection{Conversion Probabilities}

For definiteness we assume a two-flavour scenario. The oscillation
probabilities are obtained by solving the Schr\"oedinger evolution
equation of the $\nu_\mu -\nu_X$ system in the Earth--matter
background (in our case, $X = \tau $ or $s$). For {\sl neutrinos} this
equation reads:

\begin{equation}
i{\mbox{d} \over \mbox{d}t}\left(\matrix{
\nu_\mu \cr\ \nu_X\cr }\right)  = 
 \left(\matrix{
 {H}_{\mu}
& {H}_{\mu X} \cr
 {H}_{\mu X} 
& {H}_X \cr}
\right)
\left(\matrix{
\nu_\mu \cr\ \nu_X \cr}\right) \,
\label{evolution1}
\end{equation}
where
\begin{eqnarray}
H_\mu & \! = &  \! 
V_\mu + \frac{\Delta m^2}{4E_\nu} \cos2 \theta \, ,\\
H_X & \!= 
& V_X -  \frac{\Delta m^2}{4E_\nu} \cos2 \theta \, ,  \\
H_{\mu X}& \!= &  - \frac{\Delta m^2}{4E_\nu} \sin2 \theta \,,
\end{eqnarray}
with the following definition of the neutrino potentials in matter
\begin{eqnarray}
\label{potential}
V_\mu=V_\tau & = &\frac{\sqrt{2}G_F \rho}{M} (-\frac{1}{2}Y_n)\,, 
\\
V_s &= & 0\, .\\
\nonumber
\end{eqnarray}
Here $G_F$ is the Fermi constant, $\rho$ is the matter density in the
Earth, $M$ is the nucleon mass, and $Y_n$ is the neutron fraction. For
anti-neutrinos the signs of potentials $V_X$ should be reversed. In
the previous Eqs., $\theta$ is the mixing angle between the two mass
eigenstate neutrinos of masses $m_1$ and $m_2$. We define $\Delta
m^2=m_2^2-m_1^2$ in such a way that if $\Delta m^2>0$ $(\Delta m^2<0)$
the neutrino with largest muon-like component is heavier (lighter)
than the one with largest $X$--like component.

Since for the $\nu_\mu\to\nu_\tau$ case there is no matter effect, the
solution of Eq.(\ref{evolution1}) is straightforward and the
probability takes the well-known vacuum form
\begin{equation}
P_{\mu\mu}=1-\sin^2(2\theta)\sin^2 \left( \frac{\Delta m^2 L}{2E_\nu} 
\right). 
\label{probosc}
\end{equation}
where $L$ is the path-length traveled by neutrinos of
energy $E_\nu$. 

In the case of $\nu_\mu\to\nu_s$, the presence of the matter
potentials requires a numerical solution of the evolution equations in
order to obtain the oscillation probabilities $P_{\alpha\beta}$, which
are different for neutrinos and anti-neutrinos because of the reversal
of sign of the $V_X$'s. In our calculations, for the matter density
profile of the Earth we have used the approximate analytic
parametrization given in Ref.~\cite{lisi} of the PREM model of the
Earth \cite{PREM}.  Notice that for the $\nu_\mu\to\nu_s$ case, there
are two possibilities depending on the sign of $\Delta m^2$.  For
$\Delta m^2 > 0$ the matter effects enhance {\sl neutrino}
oscillations while depress {\sl anti-neutrino} oscillations, whereas
for the other sign ($\Delta m^2<0$) the opposite holds.

Finally, we have not considered oscillations of $\nu_\mu$'s into
electron neutrinos. This possibility is nowadays ruled out since
$\nu_\mu \to \nu_e$ oscillations cannot describe the measured angular
dependence of muon-like contained events \cite{atm98}. Moreover the
most favoured range of masses and mixings for this channel has already
been excluded by the negative results from the CHOOZ reactor
experiment \cite{chooz}.

\section{Atmospheric Neutrino Data Fits} 

In this Section, we describe the fitting procedure we employ in order
to determine the atmospheric neutrino parameters $\sin^2 (2\theta)$
and $\Delta m^2$ for the two conversion channels of interest and we
present our results.

\subsection{Statistical analysis}

We first notice that we rely on the separate use of the event numbers
(not of their ratios), paying attention to the correlations between
the sources of errors in the muon and electron predictions as well as
the correlations among the errors of the different energy data
samples.

A detailed description of the steps required in order to determine
the allowed regions of oscillation parameters was given in Ref.~\cite{atm98}.  
Following Refs.~\cite{atm98,fogli2,fogli3}, we define a $\chi^2$--function as
\begin{equation}
\chi^2 \equiv \sum_{I,J}
(N_I^{DATA}-N_I^{TH}) \cdot 
(\sigma_{DATA}^2 + \sigma_{TH}^2 )_{IJ}^{-1}\cdot 
(N_J^{DATA}-N_J^{TH}),
\label{chi2}
\end{equation}
where $I$ and $J$ stand for any combination of experimental data sets
and event-types considered, {\em i.e}, $I = (A, \alpha)$ and $J = (B,
\beta)$. The latin indexes $A, B$ stand for the different experiments
or different data samples in a given experiment. The greek indexes
denote electron--type or muon--type events, {\em i.e}, $\alpha, \beta
= e, \mu$.  In Eq.~(\ref{chi2}), $N_I^{TH}$ stands for the predicted
number of events (or for the predicted value of the flux, in the case
of upgoing muons) calculated as discussed in the previous Section,
whereas $N_I^{DATA}$ is the corresponding experimental measurement.  In
Eq.~(\ref{chi2}), $\sigma_{DATA}^2$ and $\sigma_{TH}^2$ are the error
matrices containing the experimental and theoretical errors,
respectively. They can be written as
\begin{equation}
\sigma_{IJ}^2 \equiv \sigma_\alpha(A)\, \rho_{\alpha \beta} (A,B)\,
\sigma_\beta(B),
\end{equation}
where $\rho_{\alpha \beta} (A,B)$ is the correlation matrix containing
all the correlations between the $\alpha$-like events in the $A$-type
experiment and $\beta$-like events in $B$-type experiment, whereas
$\sigma_\alpha(A)$ and $\sigma_\beta(B)$ are the errors for the number
of $\alpha$ and $\beta$-like events in $A$ and $B$ experiments,
respectively. The dimensionality of the error matrix varies depending
on the combination of experiments included in the analysis.

We have computed the correlation matrix $\rho_{\alpha \beta} (A,B)$ as
described in Ref.~\cite{fogli2}.  In the case of contained events, a
detailed discussion of the errors and correlations used in our
analysis can be found in Ref.~\cite{atm98}, which, for the sake of
clarity and completeness, we summarize in the Appendix. In the same
Appendix, we include the discussion of the errors and correlations
employed for the upgoing muon data analysis.

With all the definitions discussed above, we can calculate the
$\chi^2$ in Eq.~(\ref{chi2}) as a function of the neutrino parameters.
By minimization of the $\chi^2$ with respect to $\sin^2 (2\theta)$ and
$\Delta m^2$, we determine our best fit results, while the allowed
regions are determined by the following conditions: $\chi^2 \equiv
\chi_{\mathrm min}^2 + 4.61 \, (6.1)\, [9.21] $ for a confidence level
(CL) of $90\,(95)\,[99]$ \%, respectively.

The data sets employed in the statistical analysis are the following:
(i) Frejus\cite{Frejus}, Nusex\cite{Nusex}, IMB\cite{IMB} and
Soudan2\cite{Soudan2} data sets, which refer to low-energy, contained
events. For each experiment, the total rate for $e$-like and for
$\mu$-like events is reported. We jointly analyze these data (and
hereafter we collectively denote them as FISN); (ii) $e$-like and
$\mu$-like Kamiokande data\cite{Kamiokande}, including the sub-GeV
event rate and a 5-bin zenith-angle distribution for the multi-GeV
data; (iii) Super--Kamiokande data\cite{sk99}, again comprising
$e$-like and $\mu$-like contained events, arranged into sub-GeV and
multi-GeV samples, each of which given as a 5-bin zenith-angle
distribution; and the up--going data including the stopping muon flux
(5 bins in zenith-angle) and the through-going muon flux (10 angular
bins); (iv) MACRO\cite{MACRO} and Baksan\cite{Baksan} upgoing muons
samples, each one with 10 angular bins.

In the discussion of our results, along with presenting the results of
the separate analyses of the single data sets, we analyze various
possible combinations of data samples in order to develop an
understanding of the relevance of the different sets of data, which,
we recall, refer to events characterized by different properties.  For
instance, contained events are produced by neutrinos of relatively low
energies (below a few GeV), while the upgoing muon fluxes are
originated by neutrinos whose energies cover a much wider range, from
a few GeV to hundreds of GeV. This feature, combined also with the
angular distributions, allows one to test the energy dependence of the
oscillation probabilities in the different channels.

\subsection{Results of the analysis}

As a first result of our analysis, we show the (in)consistency of the
data with the Standard Model hypothesis.  The first column of
Table~\ref{tab:chi2} reports the values of the $\chi^2$ function in
the absence of new physics, as obtained with our prescriptions and
calculated for different combinations of atmospheric data sets. We
notice that all the data sets clearly indicate deviation from the
standard model. The global analysis, which refers to the full
combination of all data sets, gives a value of $\chi^2_{SM,
{\mathrm global}}= $ 214/(75 d.o.f) corresponding to a CL of $3 \times
10^{-15}$.  This indicates that the Standard Model can be safely ruled
out.  Instead, the $\chi^2$ for the global analysis decreases to
$\chi^2_{min}=$ 74/(73 d.o.f) (45 \% CL) when assuming the
$\nu_\mu \to  \nu_\tau$ oscillation hypothesis.
 
\begin{figure}
\centerline{\protect\hbox{\psfig{file=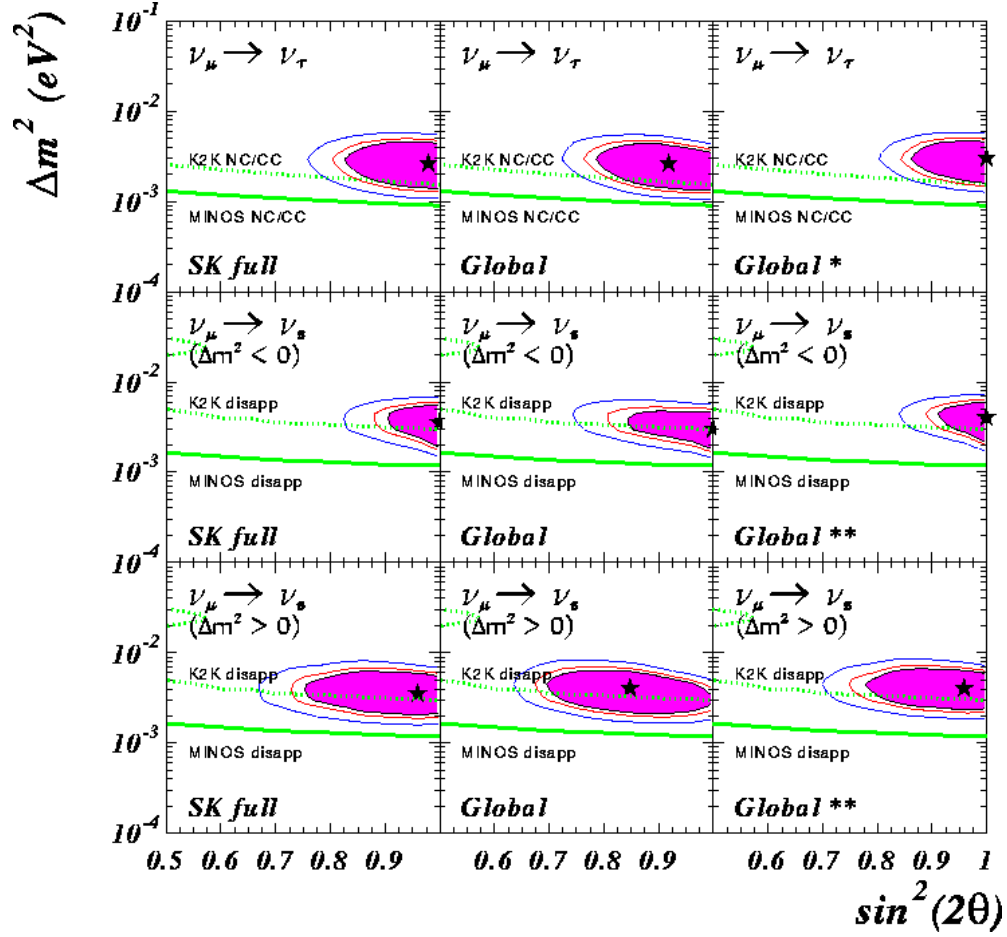,width=0.8\textwidth}}}
\vspace{15pt}
\caption{
  Allowed regions in the $\sin^2(2 \theta)$ -- $\Delta m^2$ parameter
  space for the full combination of Super--Kamiokande event data (left
  panels), the full data sample from all atmospheric neutrino
  experiments (central panels) and the combination of {\sl positive
    result} atmospheric neutrino experiments (right panels). The
  shaded areas refer to the 90\% CL, while the inner (outer) lines
  stand for 95\% (99\%) CL, respectively and the best fit points are
  indicated by a star.  With Global* we indicate that all the data
  sets are included, except for Frejus, Nusex and Baksan. Global**
  includes all the data sets, except for Frejus, Nusex, Baksan and
  MACRO. Expected sensitivities of the future long-baseline
  experiments K2K and MINOS are also displayed for each oscillation
  channel.}
\label{regionglobal}
\end{figure}

\begin{figure}
\centerline{\protect\hbox{\psfig{file=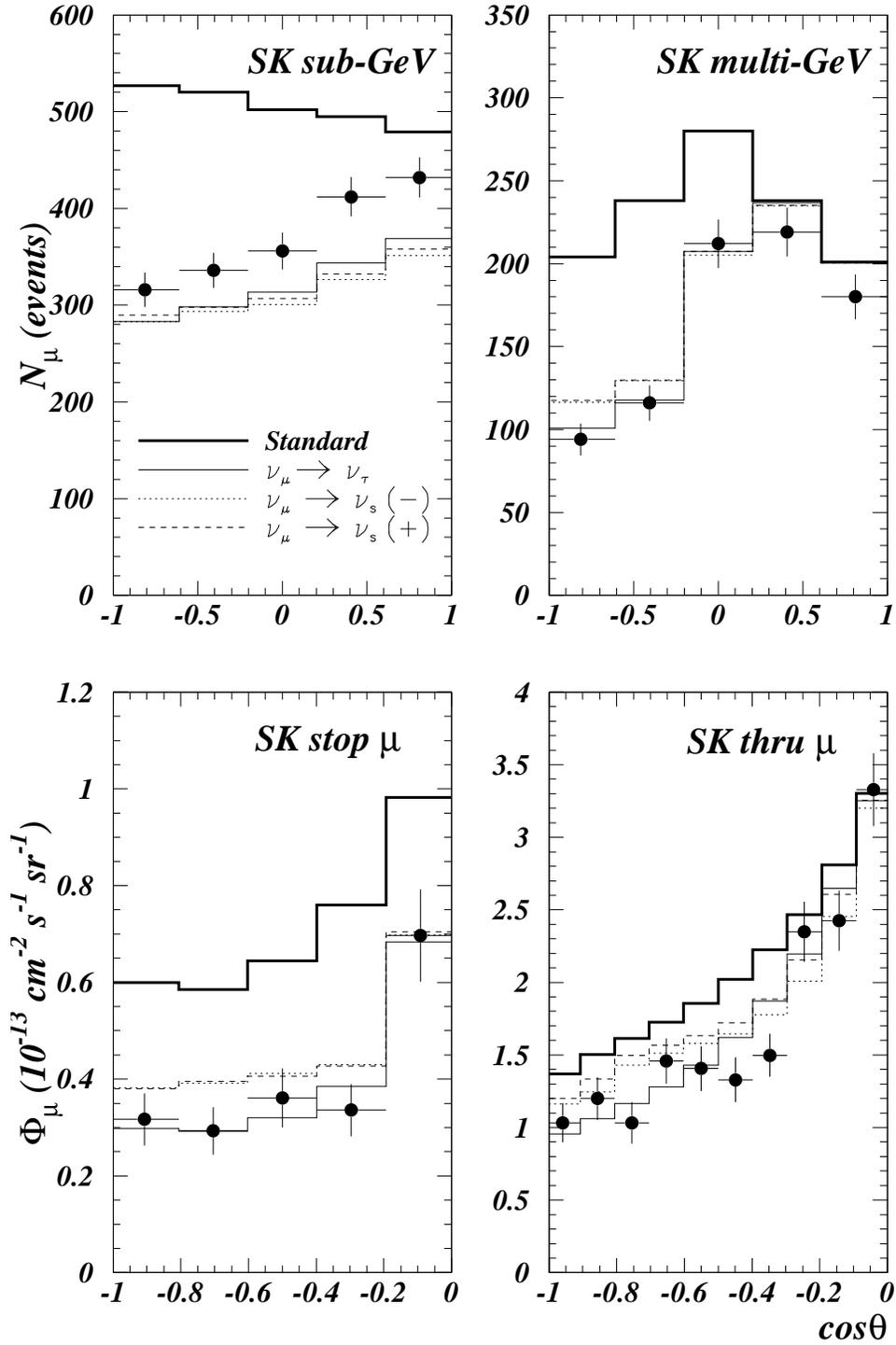,width=0.8\textwidth}}}
\vspace{15pt}
\caption{Zenith-angle distributions for the Super--Kamiokande data
sets, together with our prediction in the
absence of oscillation (thick solid line) as well as the predictions
for the best fit points obtained in the global analysis of the
data (denoted as Global** in Table~\protect\ref{tab:chi2}). 
The different lines refer to the various oscillation channels: 
$\nu_\mu \to \nu_\tau$ (thin solid line), 
$\nu_\mu \to \nu_s$ with $\Delta m^2<0$ (dotted line), and  
$\nu_\mu \to \nu_s$ with $\Delta m^2>0$ (dashed line).
The errors displayed in the experimental points are statistical.}
\label{angglobal}
\end{figure}

Table \ref{tab:chi2} reports the minimum values of $\chi^2$ and the
resulting best fit points for the various oscillation channels and
data sets considered. The corresponding allowed regions for the
oscillation parameters at 90, 95 and 99 \% CL are depicted in 
Figs.~\ref{regioncont},~\ref{regionall},~\ref{regionskup},~\ref{regionup}
and~\ref{regionglobal}. The zenith-angle distributions referring to
the Super--Kamiokande data and to the upgoing muon measurements of
MACRO and Baksan are plotted in Figs.~\ref{angcont},~\ref{angup} 
and~\ref{angglobal}.  Note that no uncertainties are shown in the plots
for the theoretical predictions, while experimental data errors are
explicitly displayed.  We now turn to the discussion of the main
results.

\subsubsection{Contained Events}

\begin{figure}
\centerline{\protect\hbox{\psfig{file=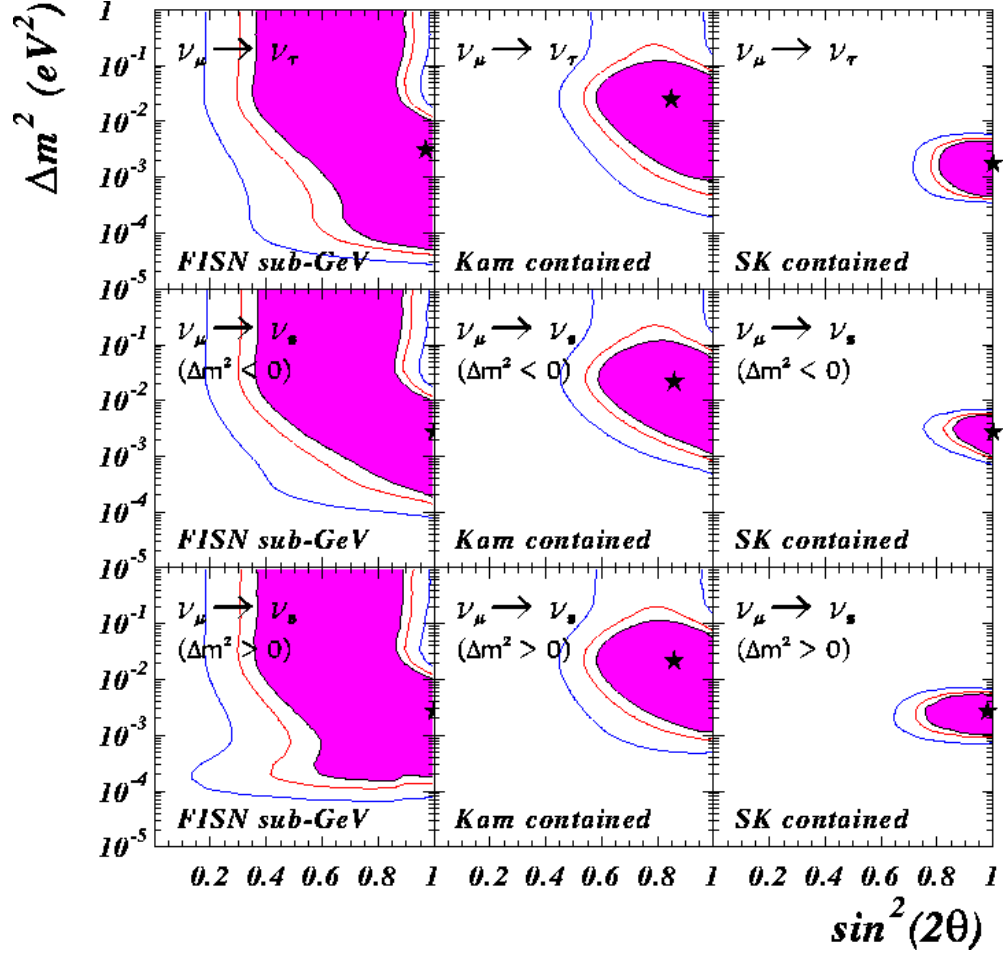,width=0.8\textwidth}}}
\vspace{15pt}
\caption{
  Allowed regions in the $\sin^2(2 \theta)$ -- $\Delta m^2$ parameter
  space for different combinations of contained events and for the
  different oscillation channels. The shaded areas refer to the 90\%
  CL, while the inner (outer) lines stand for 95\% (99\%) CL,
  respectively. FISN labels the combined rates from Frejus, IMB,
  Soudan2 and Nusex experiments.  For each panel, the best fit point
  is marked with a star.}
\label{regioncont}
\end{figure}

\begin{figure}
\centerline{\protect\hbox{\psfig{file=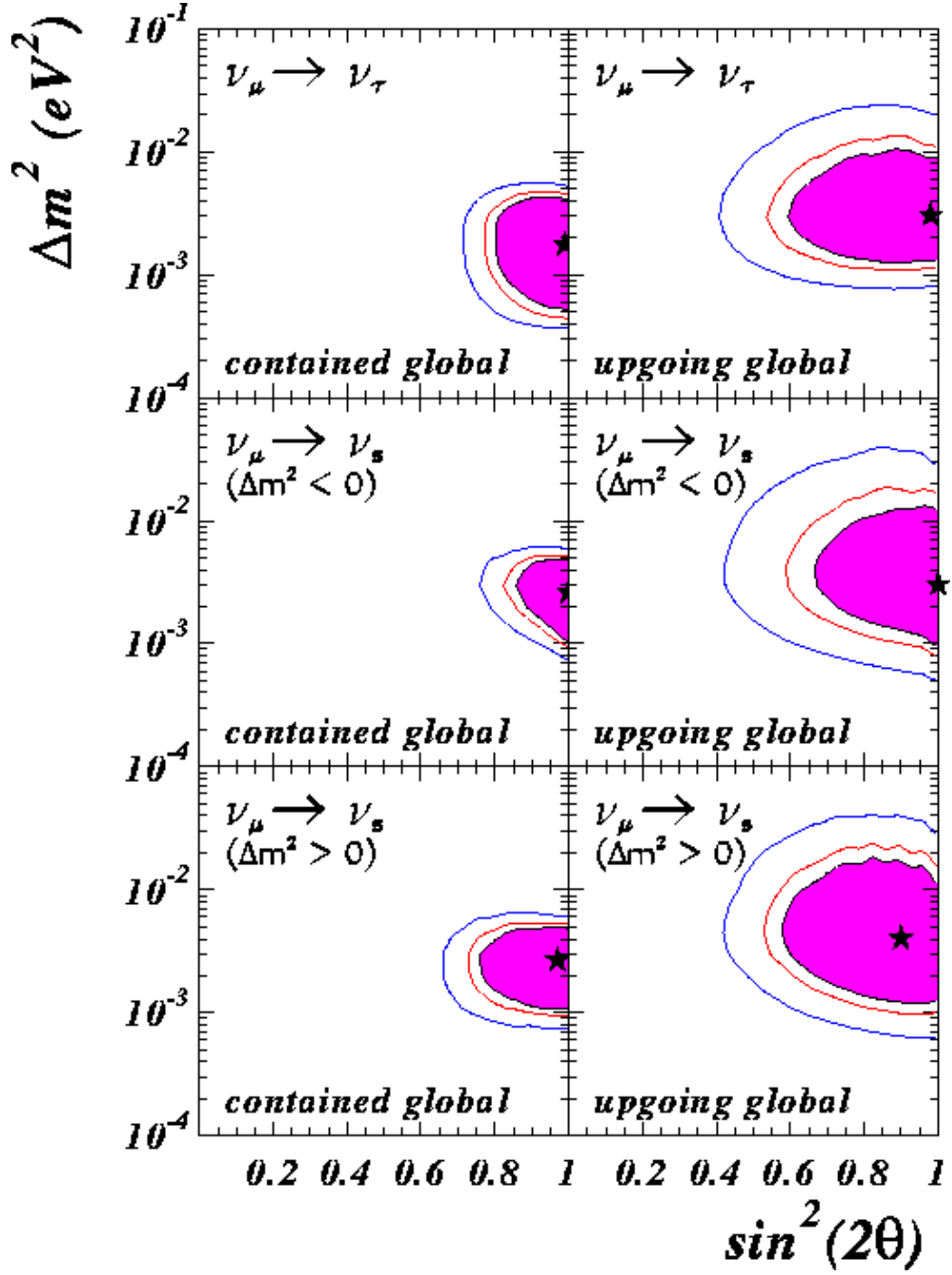,width=0.8\textwidth}}}
\vspace{15pt}
\caption{
Allowed regions in the $\sin^2(2 \theta)$ -- $\Delta m^2$ parameter space 
for the combination of all contained 
event data (left panels) and the combination of the upgoing muon data
(right panels). Notations are as in Fig. 1.}
\label{regionall}
\end{figure}

The allowed regions for contained events are displayed in
Figs.~\ref{regioncont} and ~\ref{regionall}.  In all figures the best
fit points are marked with a star and can be read from
Table~\ref{tab:chi2} together with the corresponding value of the
$\chi^2_{\mathrm min}$ for each case.

In the first column of Fig.~\ref{regioncont} we show the results for
the combination of the ``unbinned'' FISN data, {\em i.e.} the total
rates from the Frejus, IMB, Nusex and Soudan experiments. Since no
angular information in present in this case, no lower limit on the
oscillation length can be derived and the regions extend to arbitrary
large mass differences.  We notice, from Table~\ref{tab:chi2}, that
the best fit results, although definitely improved with respect to the
SM case, do not show, for each oscillation channel, a CL better than
4\%. We remind that the FISN data sample contains the Frejus and Nusex
data, which are by themselves compatible with the SM. This fact plays
a role in maintaining a relatively high $\chi^2$ even in the
oscillation cases.

The second column of Fig.~\ref{regioncont} corresponds to the
combination of contained sub-GeV (unbinned) and multi-GeV (including
the angular dependence) data from the Kamiokande experiment. In this
case, the angular distribution of the multi-GeV events plays an
important role in determining an upper limit for $\Delta m^2$, for a
given value of the mixing angle. The region which is obtained overlaps
with the previous one from the FISN data sample, and indicates a
preference for values of $\Delta m^2 \gtrsim 10^{-3}$ for all
the oscillation channels.  For the Kamiokande data, the agreement with
the oscillation hypothesis is at the level of $\sim 70$\% CL.

\begin{figure}
\centerline{\protect\hbox{\psfig{file=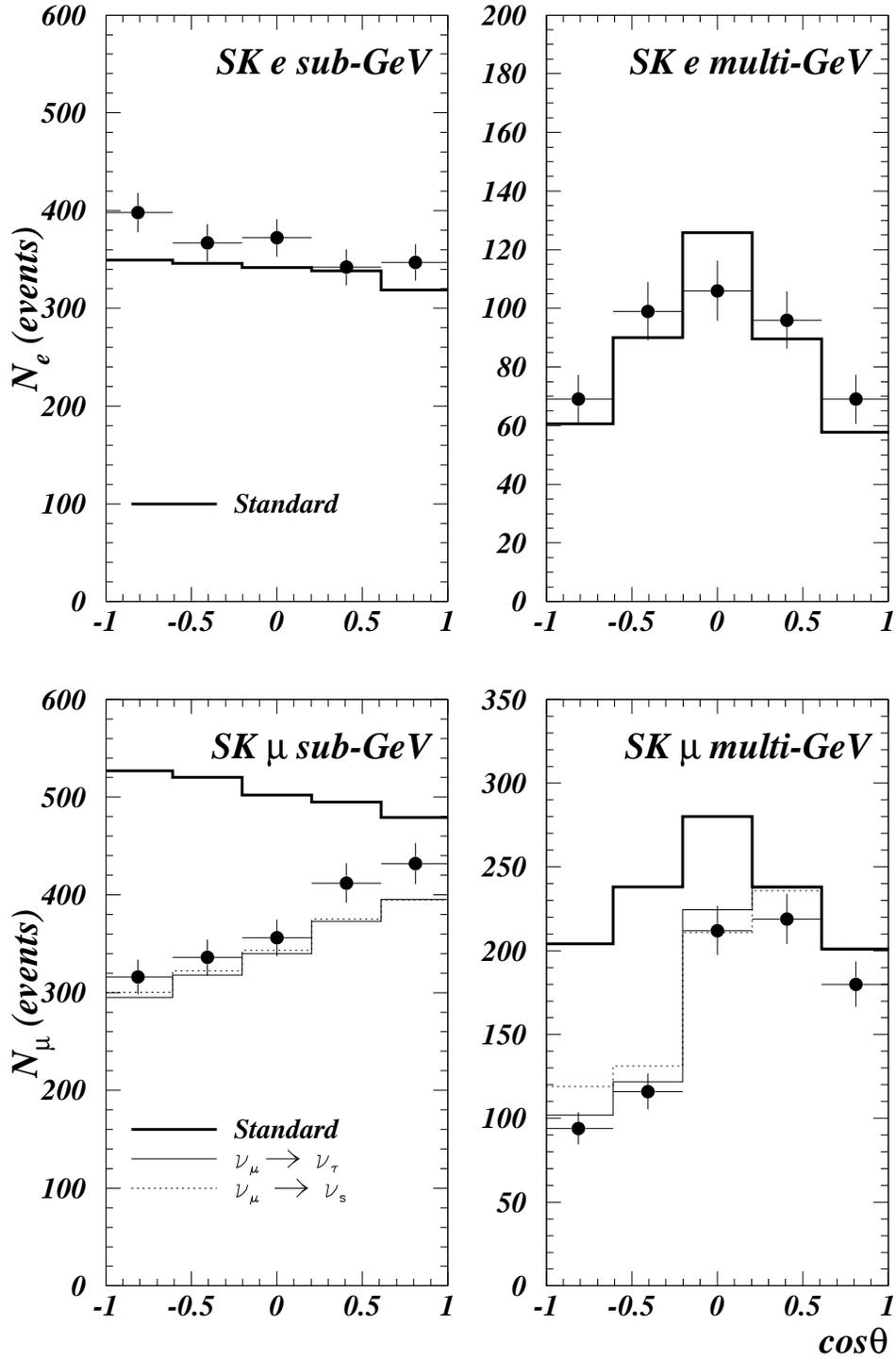,width=0.8\textwidth}}}
\vspace{15pt}
\caption{Zenith-angle distributions for the Super--Kamiokande
  electron-like and muon-like 
  sub-GeV and multi-GeV events, together with our prediction in the
  absence of oscillation (thick solid line) and the predictions for
  the best fit points for each data set in the different oscillation
  channels: $\nu_\mu \to \nu_\tau$ (thin solid line) and $\nu_\mu \to
  \nu_s$ (dotted line).  Since the best-fit point occurs for maximal
  mixing, the histograms for $\nu_\mu \to \nu_s$ are independent of
  the $\Delta m^2$ sign.  The errors displayed in the experimental
  points are statistical.}
\label{angcont} 
\end{figure}

Finally, in the third column of Fig.~\ref{regioncont} we plot the
allowed regions for the combination of the angular distributions of
the sub-GeV and multi-GeV 52 kton-yr Super--Kamiokande data. The best
fit for the $\nu_\mu \to \nu_\tau$ hypothesis has a very high
confidence level: $\chi_{\mathrm min}^2 = $ 8.9/(18 d.o.f.), which
translates into 96\% CL, substantially improved with respect to the
Kamiokande data sets. In the case of oscillation to sterile neutrinos,
the CL is 79\%, slightly higher than for the Kamiokande data alone.
The angular distributions of the Super--Kamiokande sub-GeV and
multi-GeV $\mu$-like events are shown in Fig.~\ref{angcont}. We also
show in the figure the $e$--like distributions to illustrate that they
do no show any evidence of deviation from the standard model
prediction. The agreement of the $e$--like distributions with the
expectations from the standard model has become more tight as the
Super--Kamiokande collaboration has been increasing their data sample.
This has translated into an overall improvement of the CL for the
oscillation hypothesis into channels not involving electron neutrinos.
Conversely solutions involving oscillations into electron neutrinos
have become more disfavoured \cite{fogli3,pilar}.  

From Fig.~\ref{angcont} one sees that there is a strong evidence of a
depletion in the event rates with respect to the SM expectation.  We
notice that the zenith-angle distributions obtained with the best-fit
neutrino parameters are able to reproduce the data to a high level of
agreement.

From Fig.~\ref{regioncont}, one can notice that in all the
$\nu_\mu \to  \nu_s$ channels where matter effects play a role,
the range of acceptable $\Delta m^2$ is slightly shifted towards
larger values, when compared with the $\nu_\mu \to \nu_\tau$ case.
This follows from looking at the relation between mixing {\sl in
  vacuo} and in matter. In fact, away from the resonance region,
independently of the sign of the matter potential, there is a
suppression of the mixing inside the Earth. As a result, there is a
lower cut in the allowed $\Delta m^2$ value, and this lies higher than
what is obtained in the case of the $\nu_\mu \to \nu_\tau$ channel.
Also, for the $\nu_\mu\to\nu_s$ case with $\Delta m^2 > 0$ the matter
effects enhance {\sl neutrino} oscillations while depress {\sl
  anti-neutrino} oscillations. Since atmospheric fluxes are dominantly
neutrinos, smaller mixing angle values can lead to the same $\nu_\mu$
suppression and the region extends to smaller mixing angles in the
$\Delta m^2$ region where the matter effects are important for the
relevant neutrino energies.  The opposite holds for $\Delta m^2 < 0$.
In this case the matter effects depress {\sl neutrino} oscillations,
and therefore larger mixing angles are needed to account for the
observed deficit. As a consequence, the allowed regions become smaller
(in angle) for this channel.

When comparing our results with the corresponding analysis presented
by the Super--Kamiokande Collaboration on their data sets, we find
good agreement on the allowed parameters although in general our
regions are slightly larger. We also find that, for the $\nu_\mu \to
\nu_\tau$ channel, the allowed regions for the contained events are
shifted towards slightly lower $\Delta m^2$ values with respect to
Super--Kamiokande analysis. We have traced this difference back to the
sub-GeV sample and to our use of different neutrino fluxes. Notice
that sub-GeV events are most sensitive to details in the neutrino
fluxes from the different calculations, such as the modeling of the
geo--magnetic cut--off \cite{gaisser}.  The expected angular
distribution in the absence of oscillations from the Super--Kamiokande
Monte--Carlo using Honda fluxes is slightly ``flatter'' then our
calculations for the sub--GeV event distributions. We have verified
that if we normalize our results to the Super--Kamiokande Monte--Carlo
predictions, then the allowed region we obtain is shifted to slightly
larger $\Delta m^2$ values in agreement with the Super--Kamiokande
analysis.
    
Finally, the global analysis of the combination of all the above data
sets is shown in the first column of Fig.~\ref{regionall} (notice a
change of scale between this figure and Fig.~\ref{regioncont}).  The
allowed regions, as well as the best fit points, are clearly driven by
the high statistics Super--Kamiokande data. The results show, as
expected from the above discussion, that the $\nu_\mu \to \nu_\tau$
case prefers slightly lower values of $\Delta m^2$ (from a few
10$^{-4}$ to a few 10$^{-3}$ eV$^2$) when compared to the $\nu_\mu \to
\nu_s$ cases, where values in excess of 10$^{-3}$ eV$^2$ are obtained.
In all the cases, almost-maximal mixing is statistically preferred.
Concerning the quality of the fits, as seen in Table~\ref{tab:chi2},
in the full combination the $\nu_\mu \to \nu_\tau$ channel gives a
slightly better fit ($\chi^2_{min}=37/38$) than $\nu_\mu \to \nu_s$
($\chi^2_{min}=40/38$) although the difference is not statistically
significant (51\% CL versus 38\% CL).
 
\subsubsection{Upward Going Muons}

\begin{figure}
\centerline{\protect\hbox{\psfig{file=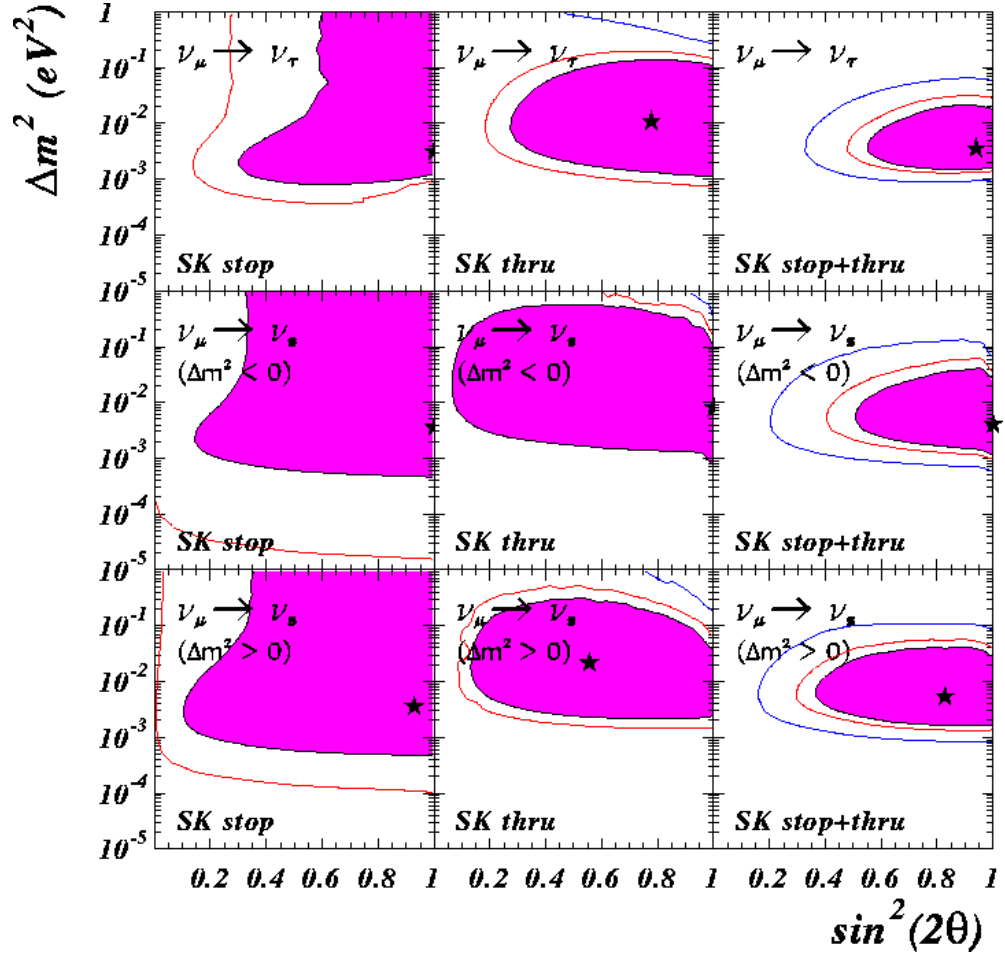,width=0.8\textwidth}}}
\vspace{15pt}
\caption{
Allowed regions in the $\sin^2(2 \theta)$ -- $\Delta m^2$ parameter space 
for the Super--Kamiokande data on upgoing muons. Notations are as in Fig. 1.}
\label{regionskup}
\end{figure}

\begin{figure}
\centerline{\protect\hbox{\psfig{file=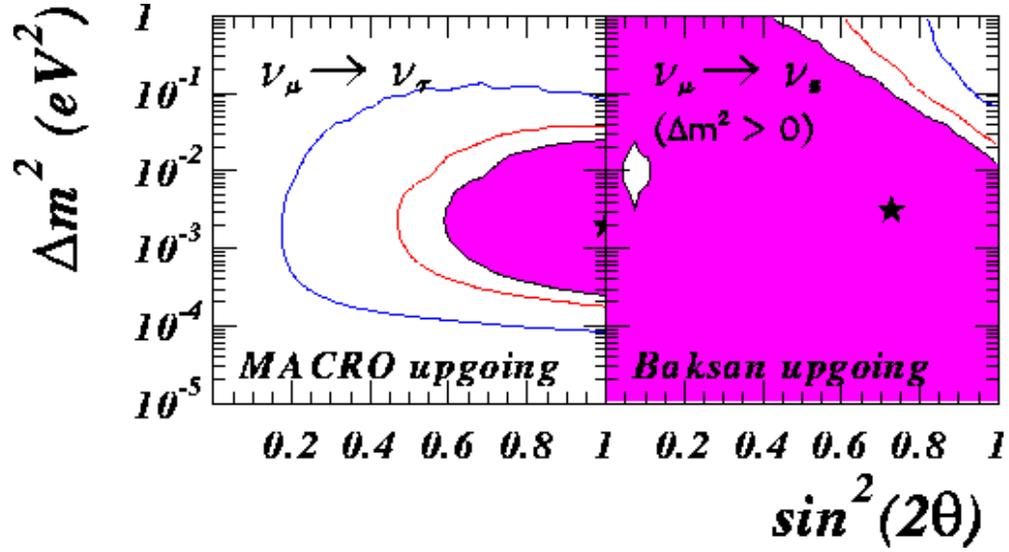,width=0.8\textwidth}}}
\vspace{15pt}
\caption{
Allowed regions in the $\sin^2(2 \theta)$ -- $\Delta m^2$ parameter space 
for the MACRO and Baksan data. The cases which provide the best
agreement with the data are presented: $\nu_\mu \to \nu_\tau$ for
MACRO and $\nu_\mu \to \nu_s$ with $\Delta m^2>0$ for Baksan.
Notations are as in Fig. 1.}
\label{regionup}
\end{figure}

Our results for the allowed regions for upgoing muon events are
displayed in Figs.~\ref{regionskup} for the Super--Kamiokande data on
stopping and through-going muons and in Fig.~\ref{regionup} for the
MACRO and Baksan experiments.  The global analysis on all the data
samples on upgoing muons is shown in the second column of Fig.~\ref{regionall}.

\begin{figure}
\centerline{\protect\hbox{\psfig{file=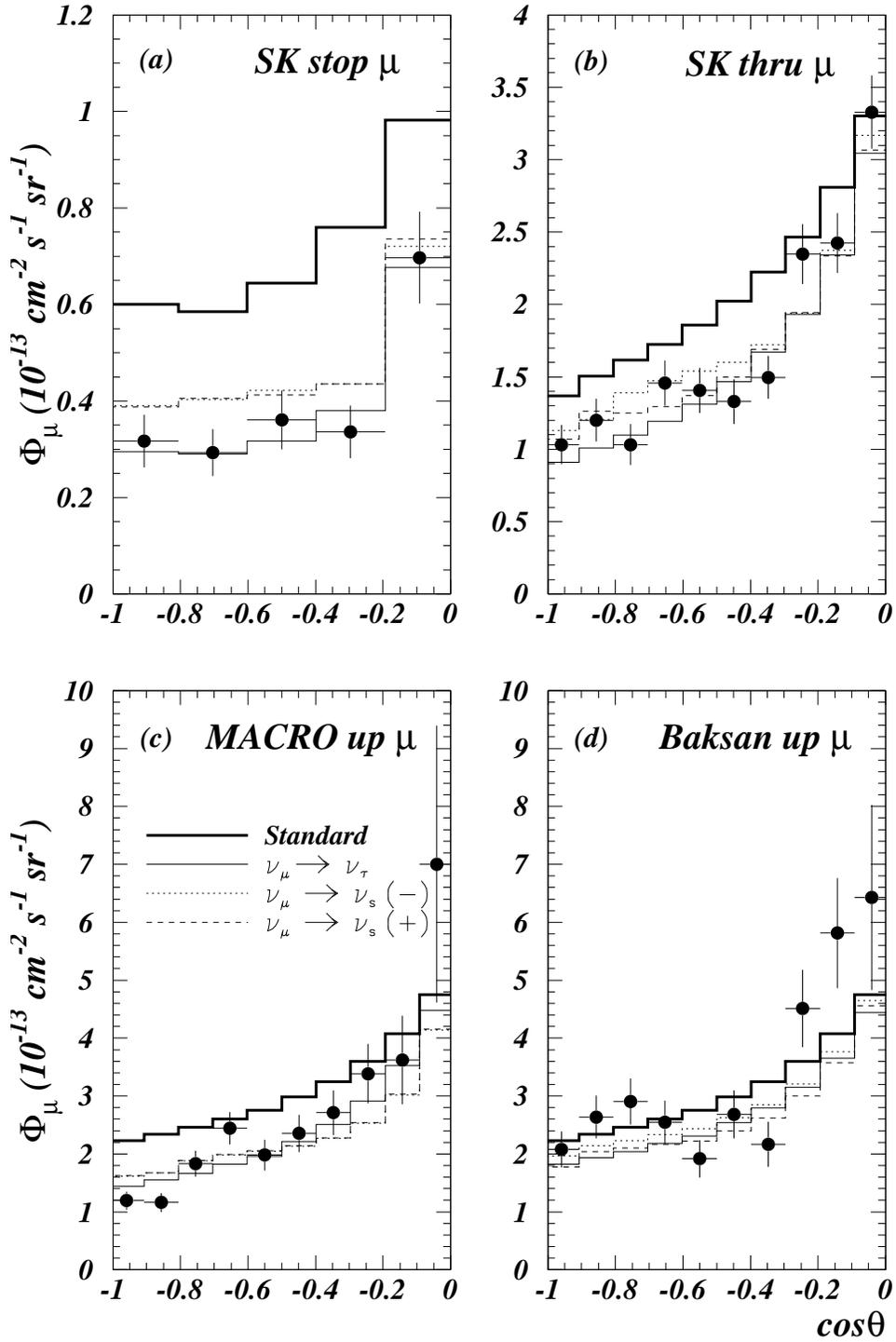,width=0.8\textwidth}}}
\vspace{15pt}
\caption{Zenith-angle distributions for upward-going muon events in 
Super--Kamiokande, MACRO and Baksan, together with our prediction in the
absence of oscillation (thick solid line) as well as the predictions
for the best fit points for each data set. The different lines refer
to: $\nu_\mu \to \nu_\tau$ (thin solid line), 
$\nu_\mu \to \nu_s$ with $\Delta m^2<0$ (dotted line), and  
$\nu_\mu \to \nu_s$ with $\Delta m^2>0$ (dashed line).
The errors displayed in the experimental points are statistical.}
\label{angup} 
\end{figure}

From the analysis of Super--Kamiokande data on stopping muons we find
that the $\chi^2$-function is substantially flat for $\Delta m^2$
values above 10$^{-3}$ eV$^2$ and $\sin^2 (2\theta) \gsim 0.5$ and
therefore the allowed regions are open from above, also at 90 \% CL.
This is a consequence of the fact that the stopping muons data sample
by its own is consistent with a global reduction of the neutrino flux
with no specific angular dependence. This feature can be observed by
comparing the angular distribution of the data points in the first
panel of Fig.~\ref{angup} and the corresponding prediction in the
absence of oscillations. This behaviour is what is expected from
oscillations with short oscillation lengths (high $\Delta m^2$).

In contrast, the through-going Super--Kamiokande muon sample indicates
a somewhat angular-dependent suppression.  The second panel of
Fig.~\ref{angup} shows that the reduction is clearly larger for
vertically-coming muons than for those arriving from the horizon.  As
a consequence, the corresponding allowed regions turn out to be closed
from above, {\em i.e.} for large $\Delta m^2$. This is plotted in the
second column of Fig.~\ref{regionskup}, whose comparison with the first
column shows the general agreement between the stopping and
through--going muons data samples. This is an interesting property in
favour of the neutrino oscillation hypothesis, since, as noted before,
stopping and through--going muon events originate from very different
parent--neutrino energies.

Finally, the combination of both stopping and thru-going muon events
in Super--Kamiokande is shown in the third column of
Fig.~\ref{regionskup}.  We find a very good agreement between our
results and the corresponding ones given by the Super--Kamiokande
Collaboration. The best fit points in ours and the Super--Kamiokande
analyses are very similar, although also in this case our regions are
slightly larger. From the angular distributions of Fig.~(\ref{angup}),
we can observe that due to matter effects the distribution for upgoing
muons in the case of $\nu_\mu \to \nu_s$ are flatter than for $\nu_\mu
\to \nu_\tau$ \cite{lipari,lipari1}.  Since the data show a somehow
steeper angular dependence, a better description in terms of $\nu_\mu
\to \nu_\tau$ oscillations is found.  From Table~\ref{tab:chi2}, we
see that the upward going muon fit in this channel is indeed better,
although the statistical preference of this channel over the sterile
case (45\% CL for $\nu_\mu \to \nu_\tau$ as compared to about 30\% CL
for $\nu_\mu \to \nu_s$) is not overwhelming.

In the case of both MACRO and Baksan experiments we generally find a
less significant statistical agreement between the data and the
theoretical evaluations.  Both data sets are clearly inconsistent with
the SM predictions, as indicated by their high $\chi^2_{SM}$ values in
Table \ref{tab:chi2}: the CL are $7 \times 10^{-4}$ for MACRO and $5
\times 10^{-3}$ for Baksan. However, the quality of the fits does not
strongly improve when interpreted in terms of neutrino oscillations.
We find that the best agreement occurs for the oscillation in the
$\nu_\mu \to \nu_\tau$ channel for the MACRO experiment. In this case
we obtain a 5\% CL and an allowed region in the parameter space, which
is shown in the first panel of Fig.~\ref{regionup}.  This result is
fully consistent with the analysis performed by the MACRO
Collaboration on their data set~\cite{MACRO}.  We notice that, as can
be seen from the angular distribution of Fig.~\ref{angup}, the flux in
the fourth bin ($-0.7 < \cos\theta_Z < -0.6$) is at least 2--$\sigma$
above the expectation in the presence of oscillations. When removing
this point from the statistical analysis, the quality of the fit for
the $\nu_\mu \to \nu_\tau$ oscillation channel improves to 19\% CL,
while affecting very little the allowed region. 
In contrast, for the sterile cases, we encounter a very low
statistical confidence (about 1\% CL), even after removing the fourth
bin data point. As a result the corresponding allowed oscillation
parameter regions are not very meaningful, since the best fit point
has a large $\chi^2_{\mathrm min}$ value. For this reason we have not
reproduced them here.  We only comment that the $\chi^2$ is very flat
in the neutrino oscillation parameters, without a clear indication of
a statistically preferred region.  In the $\Delta m^2 > 0$ case, two
almost degenerate minima are found, one for small and one for large
mixing angle, as can be seen  in Table~\ref{tab:chi2}).

For Baksan we find no clear preference for the oscillation hypothesis
with respect to the Standard Model case, although in the sterile
channels the best fits with oscillation turn out to be slightly better
than for the no-oscillation case. We reproduce in Fig.~\ref{regionup}
the allowed region for the $\nu_\mu \to \nu_s$ ($\Delta m^2 > 0$)
case, which is the one with lowest $\chi^2$. In the case of 
$\nu_\mu \to \nu_\tau$ and $\nu_\mu \to \nu_s$ ($\Delta m^2 < 0$),
the allowed regions are very similar to the one depicted in Fig.~\ref{regionup}.
We wish to warn that, due to the low statistical significance of 
the best fit results, these regions should be taken only as indicative.

To conclude this Section, we show in the right column of
Fig.~\ref{regionall} the result of our analysis for the combination of
all the data on upgoing muons discussed above. In the case of $\nu_\mu
\to \nu_\tau$, one notices that the Super--Kamiokande and MACRO
results give consistent and similar allowed regions, while Baksan
gives compatible but not strongly constraining results. As a
consequence, the combined analysis gives a region which is
intermediate to the Super--Kamiokande and MACRO ones.  The best fits
of the combined upgoing neutrino analysis have a low CL (around 1\%
for all the oscillation channels), a result which is mainly driven by
the high $\chi^2_{\mathrm min}$ values of MACRO and Baksan (we recall
that Super--Kamiokande upgoing-muon data alone indicate a preference
for neutrino oscillation at a level always better than 25\%).
However, the global analysis of the upgoing muons data disfavours the
Standard Model at the $3 \times 10^{-5}$ level.

\subsubsection{Global Analysis}

Let us now move to the discussion of the comparison and combination of
contained events with upgoing muons fluxes.  We observe from
Fig.~\ref{regionall} that the allowed regions obtained from both types
of analyses are fully consistent between themselves for all the
oscillation channels. For the cases of $\nu_\mu \to \nu_s$ the allowed
region for the contained events lies always inside the corresponding
regions allowed by the upgoing muon analysis. For the $\nu_\mu \to
\nu_\tau$ channel, we find that the region for contained events
extends to lower values of $\Delta m^2$, when compared to the region
for upgoing muons.

In figure~\ref{regionglobal} we display the allowed regions after
combining together contained and upward going muon data. In the first
column we show the results when only including the Super--Kamiokande
data.  The second column shows the corresponding results when data
from all experiments are included, while in the third column we show
the allowed regions when only the results from experiments observing
some evidence of neutrino oscillation are included.  The general
behaviour is that when including the results from all experiments the
regions become slightly larger than those obtained from the analysis
of the Super--Kamiokande data alone. On the other hand, as expected,
the results become more restrictive when only the data from
experiments observing some evidence for oscillations are included.  In
Table~\ref{tab:chi2} we list the values of the best fit points for the
various cases.

Our results from the combined analysis show that the channel $\nu_\mu
\to \nu_\tau$ gives a better fit to the data than oscillations into
sterile neutrinos. The difference for the global analysis is
$\chi^2_{min} = $ 74/(73 d.o.f.) (45 \% CL) for the active case versus
$\chi^2_{min} = $ 90/(73 d.o.f.) (8.5 \% CL) for oscillations into
sterile neutrinos with $\Delta m^2 <0$ and $\chi^2_{min} = $ 86/(73
d.o.f.) (14 \% CL) for $\nu_\mu \to \nu_s$ with $\Delta m^2 >0$.  This
difference in the quality of the description is still maintained when
some of the negative-result experiments are excluded from the analysis
(by ``negative-result experiments'' we mean experiments which, from
our statistical analysis, do not give a clear evidence of neutrino
oscillation, {\em i.e.} have relatively high $\chi^2_{\mathrm min}$
values). All of these features can be seen in Table~\ref{tab:chi2}.
When removing from the analysis the Frejus, Nusex and Baksan data
points, we do not obtain an improvement for the sterile cases, while
for the active case the CL is increased to 58 \% ($\chi^2_{min} = $
58/(61 d.o.f.)). When also MACRO is removed, we obtain higher CL also
for the sterile cases, but the $\nu_\mu \to \nu_\tau$ hypothesis
remains as the best option: $\chi^2_{min} = $ 41/(51 d.o.f.) (84 \%
CL) for the active case, $\chi^2_{min} = $ 51/(51 d.o.f.) (47 \% CL)
for $\nu_\mu \to \nu_s$ with $\Delta m^2 <0$ and $\chi^2_{min} = $
50/(51 d.o.f.)  (51 \% CL) for $\nu_\mu \to \nu_s$ with $\Delta m^2
>0$.

In conclusion, we find that the quality of the global description is
better for the $\nu_\mu \to \nu_\tau$ channel although oscillations
into $\nu_\mu \to \nu_s$ cannot be statistically ruled out on the
basis of the global fit to the full set of observables. In
Fig.~\ref{angglobal} we show the zenith-angle distributions for the
Super--Kamiokande data sets, calculated for the best--fit points
obtained in the global analysis of the data, when only ``positive
results'' experiments are considered. From the figure, we notice that
the $\nu_\mu \to \nu_s$ cases have more difficulties in reproducing
the distribution of the data points because for $\nu_\mu \to \nu_s$
the survival probability of muon--neutrinos has a less steep angular
behaviour as compared with the $\nu_\mu \to \nu_\tau$ case due to
the Earth matter effects present in the $\nu_\mu \to \nu_s$ channels.

\section{Discussion}

In this paper we have performed a global analysis of the atmospheric
neutrino data in terms of neutrino oscillations. We have compared the
relative statistical relevance of the active-active and active-sterile
channels as potential explanations of the atmospheric neutrino
anomaly.  In the analysis we have included, for the contained events,
the latest data from Super--Kamiokande, corresponding to 52 kton-yr,
together with all other available atmospheric neutrino data in the
sub-GeV and multi-GeV range. Specifically, we included the data sets
of the Frejus, Nusex, IMB, Soudan2 and Kamiokande experiments.  Our
analysis also includes the results on neutrino induced upgoing muons
from the Super--Kamiokande 52 kton-yr sample, from MACRO and from
Baksan experiments.  We have determined the best-fit neutrino
oscillation parameters and the resulting allowed regions in
$\sin^2(2\theta)$ and $\Delta m^2$ for $\nu_\mu \to \nu_X$
($X=\tau,s$) channels. For oscillations into sterile neutrinos we have
considered both positive and negative $\Delta m^2$, since the two
cases differ in the matter effect for neutrinos propagation in the
Earth.

The results of the combined analysis indicate that the channel $\nu_\mu
\to \nu_\tau$ gives a better fit to the data than oscillations into
sterile neutrinos: $\chi^2_{min} = $ 74/(73 d.o.f.) (45 \% CL) for the
active case versus $\chi^2_{min} = $ 90/(73 d.o.f.) (8.5 \% CL) for
oscillations into sterile neutrinos with $\Delta m^2 <0$ and
$\chi^2_{min} = $ 86/(73 d.o.f.) (14 \% CL) for $\nu_\mu \to \nu_s$
with $\Delta m^2 >0$. Since for some experiments we obtain large
$\chi^2_{min}$ values also for neutrino oscillations, values not
clearly better than the analysis for the standard model case, we
decided to perform additional analyses by removing these
``negative-result experiments''.  When excluding Frejus, Nusex and
Baksan data points, we do not obtain an improvement for the sterile
cases, while for the active case the CL is increased to 58 \%
($\chi^2_{min} = $ 58/(61 d.o.f.)). When also MACRO is removed, we
obtain higher CL also for the sterile cases, but the $\nu_\mu \to
\nu_\tau$ hypothesis remains as the best option: $\chi^2_{min} = $
41/(51 d.o.f.) (84 \% CL) for the active case, $\chi^2_{min} = $
51/(51 d.o.f.) (47 \% CL) for $\nu_\mu \to \nu_s$ with $\Delta m^2 <0$
and $\chi^2_{min} = $ 50/(51 d.o.f.)  (51 \% CL) for $\nu_\mu \to
\nu_s$ with $\Delta m^2 >0$. We thus conclude that the quality of the
global description of the atmospheric neutrino data in terms of
neutrino oscillation is better for the $\nu_\mu \to \nu_\tau$ channel,
although oscillations into $\nu_\mu \to \nu_s$ cannot be statistically
ruled out, on the basis of the global fit to the full set of
observables.

We have also presented a sample of predicted zenith-angle
distributions for the best-fit points corresponding to the various
oscillation channels.  Specifically, we showed the angular
distribution corresponding to the four Super--Kamiokande data sets
(sub-GeV, multi-GeV, stopping muons and through-going muons) and the
angular distributions for upgoing muons at MACRO and Baksan.  By using
the zenith-angle distribution expected for contained events and
up-going muon data at Super-Kamiokande in the presence of
oscillations, we have compared the relative goodness of the three
possible oscillation channels. This allowed one to understand why the
$\nu_\mu \to \nu_\tau$ channel gives a better description than the
$\nu_\mu \to \nu_s$ cases. On the one hand, due to matter effects, one
finds that for the sterile case the up--down asymmetry in the
multi-GeV sample is slightly smaller than observed.  Moreover, also
due to matter effects, the upgoing-muon distributions in the case of
$\nu_\mu \to \nu_s$ are flatter than for $\nu_\mu \to \nu_\tau$, while
the data show a slightly steeper angular dependence which can be
better described by $\nu_\mu \to \nu_\tau$.

To conclude, we compare our determinations of the allowed neutrino
oscillation parameters from the analyses of the atmospheric neutrino
data with the expected sensitivities of future long-baseline
experiments such as K2K\cite{k2k} and MINOS\cite{minos} 
(Fig.~\ref{regionglobal}).  We notice
that, for the oscillations $\nu_\mu \to \nu_\tau$, K2K can cover most
of the 90\% CL allowed region while the MINOS test of NC/CC is
sensitive to the complete 99\% CL region of oscillation parameters.
For oscillations into sterile neutrinos, a situation where only a CC
disappearance test can be performed at long baseline experiments, K2K
can cover a portion of the region allowed at 90\% CL, while the full
99\% CL allowed regions are completely accessible to the MINOS
experiment.

\section*{Appendix: Statistical Analysis}

In this Appendix, we discuss the errors and correlations employed in 
our analysis. 

\subsection{Errors}

Data errors contain the experimental statistical and systematic errors
as well as the uncertainties arising from event mis-identification, as
quoted by the experimental Collaborations.  In the theoretical
calculations of event rates and upgoing muon fluxes, we take into
account the uncertainty in the atmospheric neutrino flux and the
uncertainties in the charged-current neutrino cross-sections
\cite{fogli2}.  We also include, the Monte Carlo (MC) statistical
errors estimated by the experimental Collaborations with the
simulations of their detectors.  This uncertainty depends on the
number of simulated Monte--Carlo events.
 
The flux uncertainty is taken to be $30\%$ at lower energies, relevant
for contained events, and $20 \%$ at higher energies, characteristic
of upward going muons.  Nuclear cross-section uncertainties are taken
to be 10\% for all the contained event samples, except for Soudan2 for
which we used the values $7.5 \%$ and $6.4 \%$ for $e$-like and
$\mu$-like events, respectively \cite{Soudan}.  For stopping and
through-going muons, as cross-section uncertainties we use 11.4\% and
14.1 \%, respectively \cite{skup}.  The Monte--Carlo statistical
errors are estimated from the simulated exposure, as given by the
experimental Collaborations, under the assumption that the $e$- and
$\mu$-like contained events follow a binomial distribution.

\subsection{Correlations}

There is a large number of correlations, both from experimental and
from theoretical sources.  In our analysis, data errors between different 
experiments are assumed to be uncorrelated, {\em i.e.}
\begin{eqnarray}
\rho_{\alpha\alpha}^{DATA}(A,A) = 1&  \ (\alpha = e, \mu) 
\ \ & \mbox{for\  all data samples}\ \\
\rho_{\alpha\beta}^{DATA}(A,B) = 0&  \ (\alpha, \beta = e, \mu)
\ \  &\mbox{if}\ \ A \neq  \  B, \nonumber
\end{eqnarray}
while the correlations in the theoretical quantities referring to
different experiments (i.e., for $A \neq B$) are assumed to arise from
the correlations amongst the theoretical errors for the neutrino flux
normalization and amongst the theoretical uncertainties of the neutrino
interaction cross section.

In order to determine the different entries, we have classified the
experiments in four samples:

\vspace{10pt}
\begin{tabular}{lll}
$\bullet$ & {\em sub-GeV} (SG):    & Frejus, IMB, Nusex, Soudan2, Kam sub-GeV \\
 & & and Super--Kam sub-GeV \\
$\bullet$ & {\em multi-GeV} (MG):  & Kam multi-GeV and Super--Kam multi-GeV \\
$\bullet$ & {\em stopping muons} (STOP):   & Super--Kam \\
$\bullet$ & {\em thru-going muons} (THRU): &  Super--Kam, MACRO and Soudan.
\end{tabular}
\vspace{10pt}

We then estimate the correlations  as follows, 
\begin{displaymath}
\rho_{\alpha\beta}^{TH} (A,B) 
= \rho_{\alpha\beta}^{flux}(A,B)
\times \frac{ \sigma_\alpha^{flux}(A) 
\sigma_\beta^{flux}(B) }
{\sigma_\alpha^{TH}(A)\ \sigma_\beta^{TH}(B)}
+\rho_{\alpha\beta}^{cross}(A,B)\times \frac{ \sigma_\alpha^{cross}(A) 
\sigma_\beta^{cross}(B) }
{\sigma_\alpha^{TH}(A)\ \sigma_\beta^{TH}(B)}
\end{displaymath}
where the correlation $\rho_{\alpha\beta}^{flux}$ and $\rho_{\alpha\beta}^{cross}$ 
can be computed from the allowed uncertainties on the corresponding 
ratios by means of:
\begin{eqnarray}
\sigma^2_{Rf} (A:\alpha,B:\beta) 
\equiv &
\sigma^2\left(\frac{\Phi_\alpha (A)}{\Phi_\beta (B)}\right)
& = 
\sigma^{2\; flux}_\alpha (A) + 
\sigma^{2\; flux}_\beta (B) - 2\rho^{flux}_{\alpha\beta}(A,B) 
\sigma_\alpha^{flux}(A)
\sigma_\beta^{flux}(B) \nonumber\\
\sigma^2_{Rc}  (A:\alpha,B:\beta)
\equiv & 
\sigma^2\left(\frac{\sigma^{CC}_\alpha (A)}{\sigma^{CC}_\beta (B)}\right)
& =  
\sigma^{2\; cross}_\alpha (A) + 
\sigma^{2\; cross}_\beta (B) - 2\rho^{cross}_{\alpha\beta}(A,B) 
\sigma_\alpha^{cross} (A)
\sigma_\beta^{cross}(B) \nonumber
\end{eqnarray}

We assume that the errors in the above defined ratios of the neutrino fluxes
and the neutrino cross-sections between the different experiments 
arise from three sources: the flavour dependence, the energy dependence 
and the angular dependence \cite{fogli3}. 
\begin{eqnarray}
\sigma^2_{Rf} (A:\alpha,B:\beta) &=& \sigma^2_{Rf,flav} (A:\alpha,B:\beta)
+\sigma^2_{Rf,en} +(A:\alpha,B:\beta)+
\sigma^2_{ang} (A:\alpha,B:\beta) \nonumber \\
\sigma^2_{Rc} (A:\alpha,B:\beta) &=& \sigma^2_{Rc,flav} (A:\alpha,B:\beta)
+\sigma^2_{Rc,en} (A:\alpha,B:\beta) +
\sigma^2_{ang}(A:\alpha,B:\beta) \nonumber
\end{eqnarray}

For sub-GeV data samples, there are additional sources of
uncertainties with respect to the other data sets. For instance,
the form of the geo--magnetic
cut--off used in the neutrino flux, the nuclear modeling
for the neutrino interaction cross section, the neutrino
production in the atmosphere. These additional uncertainties 
allow a large variation of the 
expectations for sub-GeV experiments, without affecting the predictions
for higher energy events. Thus we neglect the correlations in the 
theoretical errors between sub-GeV events and any of the higher energy 
samples. Moreover the different experimental collaborations  use different 
calculations for the neutrino interaction cross section, thus we neglect
correlations between the rtheoretical errors of the interaction cross
sections for different detectors.
For all other cases we use the following values:

\begin{tabular}{ll}
$\sigma^2_{Rf,flav} (A:\alpha,B:\beta)= 0$
  & \hspace{10pt} for $\alpha=\beta$ \\
$\sigma^2_{Rf,flav} (A:e,B:\mu)=5 \%$    
  & \hspace{10pt} for A, B unbinned (SG) \\
$\sigma^2_{Rf,flav} (A:e,B:\mu)= 10 \%$    
  & \hspace{10pt} for A, B binned (SG) or (MG)  \\
$\sigma^2_{Rf,flav} (A:\mu,B:\mu)= 10 \%$    
  & \hspace{10pt} for A in (MG) and B in (STOP)  \\
$\sigma^2_{Rf,flav} (A:\mu,B:\mu)= 14 \%$    
  & \hspace{10pt} for A in (MG) and B in (THRU)  \\
$\sigma^2_{Rf,en} (A:\alpha,B:\beta)=0$    
  & \hspace{10pt} for A, B in the same sample  \\
$\sigma^2_{Rf,en} (A:\mu,B:\mu)= 5 \%$    
  & \hspace{10pt} for A in (MG) and B in (STOP)  \\
$\sigma^2_{Rf,en} (A:\mu,B:\mu)= 10 \%$    
  & \hspace{10pt} for A in (STOP) and B in (THRU)  \\
$\sigma^2_{Rf,en} (A:\mu,B:\mu)= 11 \%$    
  & \hspace{10pt} for A in (MG) and B in (THRU)  \\
$\sigma^2_{Rc,flav} (A:\alpha,B:\beta)= 0$   
  & \hspace{10pt} for $\alpha=\beta$ \\
$\sigma^2_{Rc,flav} (A:e,B:\mu)=$ 3 to 10\%    
  & \hspace{10pt} for A, B (SG)  \\
$\sigma^2_{Rc,fl} (A:e,B:\mu)=$ 2 to 4\%    
  & \hspace{10pt} for A, B in (MG)  \\
$\sigma^2_{Rc,en} (A:\alpha,B:\beta)= 0$   
  & \hspace{10pt} for A, B in the same sample  \\
$\sigma^2_{Rc,en} (A:\alpha,B:\mu)= 5 \%$    
  & \hspace{10pt} for A in (MG) and B in (STOP)  \\
$\sigma^2_{Rc,en} (A:\alpha,B:\mu)= 7 \%$    
  & \hspace{10pt} for A in (MG) and B in (THRU)  \\
$\sigma^2_{Rc,en} (A:\mu,B:\mu)= 5 \%$    
  & \hspace{10pt} for A in (STOP) and B in (THRU)  \\
$\sigma^2_{ang}(A:\alpha,B:\beta)=5 \% |\cos(\theta_A)-\cos(\theta_B)|$ 
  & \hspace{10pt} for all angular distributions
\end{tabular}
\vskip 0.5cm
With this we get that the smallest non-vanishing  
theoretical correlation within the Super--Kamiokande samples occurs 
between the theoretical errors for most vertical multi-GeV electron bin 
and the most horizontal thru-going muon bin and it takes the value 
$\rho^{TH}=0.735$ 
while, for example, the correlation between the theoretical errors of the
two most horizontal bins of the thru-going muon sample is 
$\rho^{TH}=0.989$. 

\acknowledgments We thank Ricardo Vazquez for providing us with the
codes for the evaluation of the muon energy loss, and Todor Stanev who
provided us with his atmospheric neutrino fluxes.  It is a pleasure
also to thank Mark Messier, Teresa Montaruli and Olga Suvorova for
useful discussions. This work was supported by DGICYT under grants
PB98-0693 and PB97-1261, and by the TMR network grant ERBFMRXCT960090
of the European Union.

\begin{table}[h]
\begin{tabular}{|l|c|c|c|c|c|c|c|c|c|c|c|}
Data sets & $\chi^2_{SM}$ &\multicolumn{3}{c|} {$\sin^2(2\theta)$} 
& \multicolumn{3}{c|} {$\Delta m^2$ ($10^{-3}$ eV$^2$) } & 
\multicolumn{3}{c|} {$\chi^2_{\mathrm min}$} & d.o.f. \\
\hline
& &  $\nu_\tau$ & $\nu_s$ & $\nu_s$ & 
  $\nu_\tau$ & $\nu_s$ & $\nu_s$ & 
  $\nu_\tau$ & $\nu_s$ & $\nu_s$ & \\
& &  & $(-)$  & $(+)$ &  
  & $(-)$  & $(+)$ &  
  & $(-)$  & $(+)$ &  \\
\hline
FISN                 &27 & 0.97 & 0.99 & 1.00 & 3.0 & 2.6 & 2.8 & 13  & 13  & 13 & 6 \\
Kamioka sub-GeV      &16 & 0.82 & 0.86 & 0.95 & 400 & 12  & 7.0 & 0.02 & 0.02 &0.02 & 0 \\
Kamioka multi-GeV    &19 & 0.95 & 0.95 & 0.93 & 23  & 25  & 25  & 6.8  & 6.8  &6.7 & 8 \\
{\it Kamioka combined}&35& 0.85 & 0.86 & 0.86 & 25  & 22  & 21  & 7.3  & 7.2 & 7.0 & 10\\
\hline
SK sub--GeV        & 27   & 1.0  & 1.0  & 1.0  & 1.3 & 1.3 & 1.3 & 2.4 & 2.7 &2.7 & 8 \\
SK multi--GeV
      & 42   & 0.98 & 1.0  & 0.99 & 1.7 & 3.5 & 3.5 & 6.3 & 9.0 & 8.9 & 8 \\
{\it SK contained}& 69   & 1.0  & 1.0  & 0.98 & 1.7 & 2.7 & 2.6 & 8.9 & 13 &13&18\\
SK stop $\mu$     & 8.4  & 1.0  & 1.0  & 0.93 & 3.0 & 3.5 & 3.5 & 1.3 & 2.4 &2.3 & 3 \\
SK thru $\mu$     & 19   & 0.78 & 1.0  & 0.53 & 11  & 8.1 & 21  & 10 &13 &10&  8 \\
{\it SK upgoing}  & 30   & 0.94 & 1.0  & 0.86 & 3.5 & 4.0 & 5.3 & 13 &16 &15&13 \\
{\it SK combined}     & 122  & 0.98 & 1.0  & 0.93 & 2.6 & 3.5 & 3.5 & 24& 33&32& 33 \\
\hline
MACRO thru $\mu$  & 27   & 1.0 & 1.0 & 0.07 &2.0 & 8.1& 12 & 16 & 24 &23& 8 \\
MACRO thru $\mu$  & 27    &     &     & 1.0  &     & & 7.9 &      & &24 & 8 \\
Baksan thru $\mu$ & 22   & 0.41 & 0.66 & 0.73 & 6.1 & 2.0  & 3.0 & 21 &19& 17&8\\
\hline
{\it Contained comb} & 103 & 0.99 &1.0 & 0.97 & 1.7 &2.7&2.6&37&40&40 & 38 \\
{\it Upgoing $\mu$ comb} & 76 & 0.99 &1.0& 0.97 &3.0&3.0&4.6&56 &60 &59& 33 \\
\hline
{\bf Global}   & {\bf 214} & {\bf 0.92} & {\bf 1.0} &  {\bf 0.85} 
& {\bf 2.6 } & {\bf 3.0} & {\bf 4.0}  
&{\bf 74} & {\bf 90} & {\bf 86}   & {\bf 73}\\
{\bf Global*} & {\bf 182} & {\bf 1.0} & {\bf 1.0} 
&{\bf 0.95}    & {\bf 3.0 } & {\bf 3.0} & {\bf 4.0}  
& {\bf 58} &{\bf 76} & {\bf 75}   & {\bf 61}\\
{\bf Global**} & {\bf 156 }& {\bf 0.99}  & {\bf 1.0} & {\bf 0.96}  &
{\bf 3.0}  & {\bf 4.0}& {\bf 4.0}  & {\bf 41} &  {\bf 51} & {\bf 50}  & {\bf 51}\\
\end{tabular}
\vspace{20pt}
\caption{Minimum $\chi^2$ values and best-fit points for the 
various data sets and  oscillation channels. For the 
$\nu_\mu \to\nu_s$ channels, we have  denoted by $(-)$ the 
case with $\Delta m^2<0$ and by $(+)$ the case with $\Delta m^2>0$.
{\bf Global} refers to the best-fit for the global analysis which 
includes all the data sets listed above.
{\bf Global*} includes all the data sets, except for Frejus, Nusex and 
Baksan.
{\bf Global**} includes all the data sets, except for Frejus, Nusex, 
Baksan and MACRO. For the sake of comparison we also list the 
$\chi^2_{SM}$ expected in the absence of oscillations. Notice that for 
the MACRO data set, for the  $\nu_\mu \to\nu_s$ channel 
($\Delta m^2>0$) we find two almost degenerate minima.}
\label{tab:chi2}
\end{table}


\newpage
\newpage
\newpage
\newpage
\newpage
\newpage
\newpage
\newpage

\end{document}